\documentclass{article}

\usepackage{lipsum} 
\usepackage{xcolor}
\usepackage{xspace}
\usepackage{subcaption}
\usepackage{booktabs}
\usepackage{tabularx}
\usepackage{soul}
\usepackage[framemethod=tikz]{mdframed}
\usepackage{marvosym}
\usepackage{listings} 
\usepackage{multirow}   
\usepackage{makecell}  
\usepackage[most]{tcolorbox}
\usepackage{xurl}
\usepackage{placeins}
\usepackage{enumitem}
\usepackage{microtype}
\usepackage{graphicx}
\usepackage{booktabs} 
\usepackage{amsmath}
\usepackage{hyperref}
\usepackage{fancyhdr}

\usepackage[accepted]{mlsys2026}
\usepackage{comment} 

\usepackage{tikz}
\usepackage{diagbox}
\usepackage{eso-pic}

\include{include/cmds}

\mlsystitlerunning{MLCommons Chakra: Advancing Performance Benchmarking and Co-design using Standardized Execution Traces}

\AddToShipoutPictureBG*{%
  \AtPageUpperLeft{%
    \hspace*{\paperwidth}%
    \raisebox{-68pt}{%
      \llap{%
        \href{https://www.acm.org/publications/policies/artifact-review-and-badging-current}{%
          \includegraphics[height=65pt]{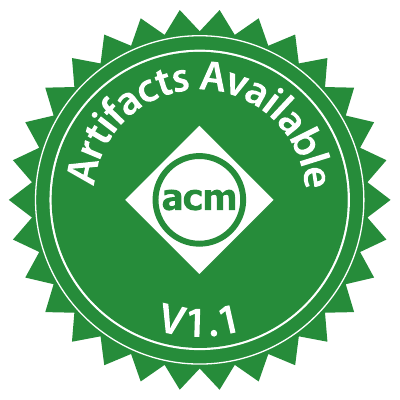}}%
        \hspace{1pt}%
        \href{https://www.acm.org/publications/policies/artifact-review-and-badging-current}{%
          \includegraphics[height=65pt]{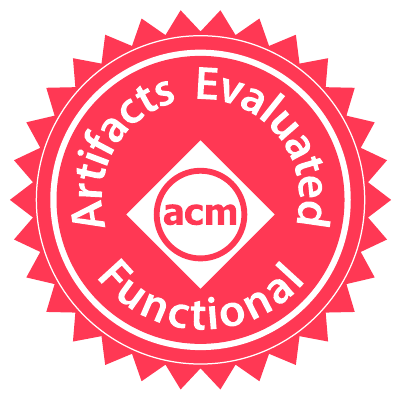}}%
        \hspace{80pt}%
      }%
    }%
  }%
}

\begin{document}
\begin{titlepage}
  \twocolumn
  \thispagestyle{empty}
\end{titlepage} 



\twocolumn[
\mlsystitle{MLCommons Chakra: Advancing Performance Benchmarking and Co-design using Standardized Execution Traces}

\mlsyssetsymbol{equal}{$*$}
\mlsyssetsymbol{work}{$\dagger$}


\begin{mlsysauthorlist}
\mlsysauthor{Srinivas Sridharan}{nvidia}
\mlsysauthor{Theodor-Adrian Badea$^\star$}{keysight}
\mlsysauthor{Andy Balogh}{keysight}
\mlsysauthor{Bradford M. Beckmann}{amd}
\mlsysauthor{Brian Coutinho}{nvidia}
\mlsysauthor{Louis Feng}{meta}
\mlsysauthor{Sheng Fu}{nvidia}
\mlsysauthor{Sanshan Gao}{nvidia}
\mlsysauthor{Mehryar Garakani}{scala}
\mlsysauthor{Taekyung Heo}{nvidia}
\mlsysauthor{David Kanter}{mlcommons}
\mlsysauthor{Josh Ladd}{nvidia}
\mlsysauthor{Ziwei Li}{gatech}
\mlsysauthor{Winston Liu}{keysight}
\mlsysauthor{Changhai Man}{gatech}
\mlsysauthor{Dan Mihailescu}{keysight}
\mlsysauthor{Spandan More}{amd}
\mlsysauthor{Joongun Park}{gatech}
\mlsysauthor{Ashwin Ramachandran}{meta}
\mlsysauthor{Vinay Ramakrishnaiah}{amd}
\mlsysauthor{Saeed Rashidi}{meta}
\mlsysauthor{Vijay Janapa Reddi}{harvard}
\mlsysauthor{Puneet Sharma}{hpe}
\mlsysauthor{Phio Tian}{nvidia}
\mlsysauthor{William Won}{amd,gatech}
\mlsysauthor{Hanjiang Wu}{gatech}
\mlsysauthor{Huan Xu}{gatech}
\mlsysauthor{Jinsun Yoo}{gatech}
\mlsysauthor{Tushar Krishna}{gatech,infravana}
\end{mlsysauthorlist}

\mlsysaffiliation{meta}{Meta}
\mlsysaffiliation{harvard}{Harvard University}
\mlsysaffiliation{keysight}{Keysight}
\mlsysaffiliation{meta}{Meta}
\mlsysaffiliation{nvidia}{NVIDIA}
\mlsysaffiliation{scala}{Scala Computing}
\mlsysaffiliation{mlcommons}{MLCommons\\}
\mlsysaffiliation{gatech}{Georgia Institute of Technology}
\mlsysaffiliation{hpe}{Hewlett Packard Enterprise}
\mlsysaffiliation{infravana}{InfraVana}
\mlsysaffiliation{amd}{AMD}

\mlsyscorrespondingauthor{\\- Srinivas Sridharan}{srinivas@mlcommons.org}
\mlsyscorrespondingauthor{\\- Tushar Krishna}{tushar@mlcommons.org}

\printAffiliationsBox




\mlsyskeywords{Machine Learning, MLSys}

\vskip 0.3in

\begin{abstract}
\label{sec:abstract}
The fast pace of artificial intelligence~(AI) innovation demands an agile methodology for observation, reproduction and optimization of distributed machine learning~(ML) workload behavior in production AI systems and enables efficient software-hardware~(SW-HW) co-design for future systems. 
We present Chakra,
an open and portable ecosystem for performance benchmarking and co-design. The core component of Chakra is an open and interoperable graph-based representation of distributed AI/ML workloads, called Chakra execution trace~(ET). These ETs represent key operations, such as compute, memory, and communication, data and control dependencies, timing, and resource constraints. Additionally, Chakra includes a complementary set of tools and capabilities to enable the collection, analysis, generation, and adoption of Chakra ETs by a broad range of simulators, emulators, and replay tools. We present analysis of Chakra ETs collected on production AI clusters and demonstrate value via real-world case studies. Chakra has been adopted by MLCommons and has active contributions and engagement across the industry, including but not limited to NVIDIA, AMD, Meta, Keysight, HPE, and Scala, to name a few.


\end{abstract}

\begin{center}
\small \textbf{Chakra Codebase:} \url{https://github.com/mlcommons/chakra}
\end{center}
]

\printNoticeOnly{\noindent Authors are listed in alphabetical order by last name, except for the corresponding authors.\\
$^\star$Theodor's name was unintentionally missed in the MLSys 2026 proceedings version of this paper.\\}
\begingroup
\endgroup




 \newcommand{\WW}[1]{\textcolor{magenta}{#1}}

\section{Introduction}
\label{sec:introduction}

As artificial intelligence~(AI) models continue to scale at an unprecedented rate in terms of size and capability, and large language models~(LLMs) continue unlocking novel serving use cases, the design and deployment of the platforms on which AI training and inference is done of paramount importance.
These AI data centers or AI supercomputers comprise of thousands to tens of thousands of customized neural processing units (NPUs) (e.g., NVIDIA Hopper/Blackwell, AMD Instinct\texttrademark{}, Google Cloud TPU, Cerebras Andromeda) connected via high-speed scale-up/scale-out interconnects. 

Unlocking performance from these platforms relies heavily on software-hardware (SW-HW) co-design across both the compute and communication stacks. Co-design is actually not a linear process, but rather a cyclic, iterative process. \autoref{fig:codesign} presents an overview of the co-design cycle that production platforms at hyperscalars go through from our experience. It involves observation and telemetry of the performance of current and emerging machine learning~(ML) workloads to identify bottlenecks, reproducing these experiments via representative benchmarks, designing next-generation platforms to address (some of) these bottlenecks, evaluating these platforms via simulation (pre-silicon) and emulation (post-silicon), implementing and testing the platform (pre-deployment), and finally deploying at scale. This cycle continues.


\insertFigure{codesign}{AI system SW-HW co-design flow.}{0.98}{-1em}{-2em}

This co-design cycle also highlights the suite of \textit{frameworks and tools that are needed for enabling efficient co-design}, spanning performance observation and analysis, reproducibility via replay, and simulators/emulators. All of these tools need to be driven by AI workloads representing current and future AI training/inference scenarios. Unfortunately, the ecosystem of these tools is extremely fragmented. Production workloads are developed and owned by AI labs and hyperscalars implemented in high-level frameworks (PyTorch/JAX), with internal observability tools, making it hard to reproduce behaviors in different environments. Simulators and emulators are a dime-a-dozen across NPU compute and networking vendors, of varying degrees of fidelity. Each of these have their own custom formats for describing workloads and the AI platform architecture. This fragmentation creates barriers to platform-agnostic analysis and co-design, and limits the opportunities for cross-platform optimizations. As a result, system-level optimizations are often limited to the boundaries of individual frameworks, restricting broader collaboration and innovation.
    
In addition, since open-source AI benchmarks, such as MLPerf, evolve slowly, there exist no efficient mechanism for workload sharing between the different players in the ecosystem. 
From our experience, hyperscalers and cloud service providers~(CSPs) are hesitant in sharing proprietary model details, and often resort to sharing spreadsheets with representative model parameters under non-disclosure agreements with select partners. However, this still makes exact workload reproduction difficult. Many hardware vendors, startups and academic teams often derive parameters from public benchmarks, which can lead to over-optimization.

Furthermore, even with access to workloads and tooling, it is  non-trivial to efficiently study optimizations on current systems and what-if scenarios for future systems. This is due to the extremely high-cost of running full-stack benchmarks as it requires (i)~heavy cross-domain expertise\footnote{Companies have dedicated teams to optimize performance before MLPerf submissions for instance.} and (ii)~access to expensive NPU clusters. Moreover, it is often hard to isolate specific HW/SW bottlenecks or  compute versus memory, versus network behavior in end-to-end runs.

The fast pace of AI innovation demands an agile methodology to create workloads and rapidly iterate through the different stages of the co-design cycle. Identifying this need, Meta led a standardization effort in early 2023~\cite{chakra_mlbench2023} called ``Chakra,"\footnote{Chakra means wheel in Sanskrit, and is inspired by the co-design cycle.} in collaboration with Georgia Institute of Technology as an academic partner, to develop an open ecosystem of methodologies and tools to enable efficient performance benchmarking, optimization, and co-design for AI platforms. 
The central idea in Chakra is that of an \textit{execution trace} (ET), a mechanism to describe distributed AI workload performance behavior over an AI platform. Analogous to instruction and memory traces~\cite{google_workload_traces}, ETs record operator dimensions for compute and communication and their dependencies while avoiding disclosure of model or dataset details. Software organizations can share ETs of internal workloads with hardware vendors, who can in turn estimate and optimize performance using proprietary hardware models and simulators. The feedback loop helps software teams select compute and network configurations for training. The availability of ETs also enables academic researchers and startups to participate in co-design for production workloads. This methodology was successfully used between Meta and NVIDIA to identify challenges overlapping compute and comms in NVIDIA V100 systems~\cite{rashidi2021enabling}.

\insertFigure{chakra_arch}{Chakra Infrastructure Overview.}{0.98}{-1em}{-2em}

As several other companies started showing an interest in this vision, Chakra was formally adopted by MLCommons \cite{chakra_mlcommons}. Today, Chakra has evolved into a 40+ member working group actively engaged in schema standardization and ecosystem (methodology and tooling) development. 
The Chakra ecosystem is already being used by several companies, spanning hyperscalers and CSPs (Meta \cite{meta_chakra_2023}, Google), AI compute/full-stack vendors (NVIDIA, AMD, HPE, Intel, Marvell), system integrators (HPE),  switch vendors (Juniper, HPE), simulation/emulation/testing tool suppliers (Keysight \cite{keysight_kai_dc_builder}, Scala Computing), and numerous academic groups and startups.
Chakra trace collection is now officially supported as an option in  both PyTorch and NVIDIA NeMo~\cite{nvidia_nemo}. For simulation, Chakra is natively supported by the open-source ASTRA-sim~\cite{astrasim-web} distributed AI simulator, providing a reference implementation to the broader community, and by several proprietary simulators, including AMD internal simulation.
Chakra has also enabled accessible benchmarking of distributed AI training on diverse platforms~\cite{go2025characterizingefficiencydistributedtraining}.

\textbf{Chakra Ecosystem.} In this paper, we present the details of the Chakra ecosystem. \autoref{fig:chakra_arch} illustrates its key components.
The Chakra ET serve as the central element, facilitating the exchange of the model behavior needed for system optimization and co-design.
Our contributions are as follows.\vspace{-1mm}
\squishlist
    \item \textbf{Schema Standardization.} We present the Chakra execution trace schema (\autoref{sec:schema}) and surrounding workflow, shaped through collaboration with multiple organizations, that aim at standardization across platforms.
    \item \textbf{End-to-end workflow.} We present the co-design workflow enabled by Chakra, introducing trace collection methodologies from upstream ML frameworks (\autoref{sec:trace_collection}) and downstream use-cases spanning trace analysis, replay, and/or simulation/emulation (\autoref{sec:use-cases}).
    \item \textbf{Case studies and artifacts.} We demonstrate our workflow via case studies spanning real trace analysis, replay benchmarks and simulation (\autoref{sec:eval}). All these artifacts are being actively released as open-source via MLCommons to enable reproduction and adoption.
\squishend

\autoref{sec:related_work} contrasts Chakra with other related efforts.

\vspace{-2mm}

\section{The Chakra Schema}
\label{sec:schema}
\vspace{-1mm}

The Chakra schema provides a standardized representation of execution traces for AI workloads. The goal is to capture execution in a form that is portable across frameworks and extensible for diverse use cases, while remaining compact and easy to consume by diverse tools.

\vspace{-1mm}
\subsection{Design Requirements}
\label{sec:chakra-schema-design-requirements}
\vspace{-1mm}

ML tasks are naturally expressed as graphs, with frameworks such as TensorFlow and PyTorch constructing internal computational graphs~\cite{computational_graphs}. In this work, we generalize this idea to \textit{execution traces}, graphs that record not only model structure but also execution details such as memory accesses, compute operations, communication, and parallelization strategies. Although ETs are a powerful abstraction, their structure and metadata differ across frameworks, limiting reuse.

The Chakra schema was designed through discussions with both academic and industrial teams to meet several requirements. First, the schema must be minimal yet extensible: only essential fields should be mandatory, but attributes should allow customization without modifying the core definition. Second, the schema must be expressive for performance modeling: it should capture computation, memory, communication, and their dependencies. Third, the schema must be portable across levels of abstraction: traces should be usable for both real system replay and projection to future system configurations.
Fourth, the schema should capture behavior at different stages of execution (\autoref{sec:trace_collection}). 

\vspace{-1mm}
\subsection{Schema Details}
\label{sec:chakra-schema-detail}
\vspace{-1mm}

The schema represents execution as a directed acyclic graph (DAG) where nodes denote operations and edges encode data and control dependencies. Instead of fixing a large set of operators, Chakra defines a small set of node categories together with an attribute mechanism for extension. This ensures portability while allowing new operators or system-specific annotations to be added without breaking compatibility. Communication is modeled explicitly as a node type alongside computation and memory, enabling consistent representation of parallelization strategies and system-level effects. By unifying these elements, the schema supports performance analysis and SW-HW co-design without exposing proprietary model or dataset details.

\begin{table}[t!]
    \centering
    \caption{Chakra node schema.}
    \resizebox{\columnwidth}{!}{
    \begin{tabular}{|c|c|c|}
    \hline
    \textbf{Field Name} & \textbf{Data Type} & \textbf{Description} \\ \hline
    id                  & uint64             & Unique node identifier \\ \hline
    name                & string             & Human-readable name \\ \hline
    type                & enum(NodeType)     & Node category (compute, memory, comms) \\ \hline
    ctrl\_deps          & repeated uint64    & Control dependencies \\ \hline
    data\_deps          & repeated uint64    & Data dependencies \\ \hline
    start\_time\_micros & uint64             & Optional start time \\ \hline
    duration\_micros    & uint64             & Optional duration \\ \hline
    inputs/outputs      & IOInfo             & Values, shapes, types \\ \hline
    attr                & repeated AttributeProto & Extensible metadata \\ \hline
    \end{tabular}
    }
    \label{tab:chakra-node-schema}
        \vspace{-1.5em}
\end{table}

Each node (\autoref{tab:chakra-node-schema}) contains a unique identifier, a name, and a type that specifies whether it represents computation, memory, or communication. Control and data dependencies define the partial order, allowing the feeder to reconstruct execution. Optional timing fields capture start and duration hints. 
Inputs and outputs are described through \texttt{IOInfo}, which records identifiers, shapes, and data types of tensors. This allows each node to be linked consistently to the tensor schema without requiring a separate table in the main text.

\begin{table}[t!]
    \centering
\caption{Chakra communication schema.}
    \resizebox{\columnwidth}{!}{
    \begin{tabular}{|c|c|}
    \hline
    \textbf{Field Name} & \textbf{Data Type} \\ \hline
    type & enum(CommType): AllReduce, AllGather, ReduceScatter\\
     & Broadcast, Point-to-Point, All2All, Barrier \\ \hline    group       & uint64 \\ \hline
    tag         & string \\ \hline
    tensor\_ids & repeated uint64 \\ \hline
    \end{tabular}
    }
    \label{tab:chakra-comm-schema}
    \vspace{-1em}
\end{table}

Compute nodes describe compute operators either on host or device. They may expose different levels of details according to proprietary needs. The \texttt{dur} field is essential, as it reflects the computation duration. Fields such as \texttt{num\_ops} and \texttt{tensor\_size} can be used to specify high-level characteristics. Moreover, attributes like \texttt{kernel\_name}, \texttt{launch\_parameter}, and \texttt{kernel\_args} can also be encoded to enable complete re-execution.

Communication nodes describe collective and point-to-point operations. 
The \texttt{type} field specifies the communication primitive. 
The \texttt{group} field specifies the set of ranks involved in the operation, 
and the optional \texttt{tag} helps distinguish concurrent operations. 
\texttt{tensor\_ids} links the communication to tensor objects 
defined in the tensor schema.

To support collective communication and parallelism across devices, Chakra uses a concept similar to \emph{process groups} in PyTorch. In PyTorch, a process group defines a set of ranks that participate in a collective operation such as \texttt{all\_reduce}, \texttt{all\_gather}, or \texttt{reduce\_scatter}. Chakra encodes the same idea by attaching a group identifier to communication nodes. This makes it possible to represent communication patterns over arbitrary subsets of NPUs. In practice, this allows traces to describe combinations of tensor, pipeline, data, and expert parallelism without altering the schema.
Through process groups, Chakra can represent complex parallelization and communication strategies in a uniform way, while keeping the schema itself minimal and device traces independent.




\begin{table}[t!]
\centering
\begin{minipage}{0.26\textwidth}
    \centering
    \caption{Tensor schema.}
    \resizebox{\linewidth}{!}{
    \begin{tabular}{|c|c|}
    \hline
    \textbf{Field Name} & \textbf{Data Type} \\ \hline
    id      & uint64 \\ \hline
    storage\_id & uint64 \\ \hline
    storage\_offset & uint64 \\ \hline
    shape   & repeated int64 \\ \hline
    stride  & repeated int64 \\ \hline
    dtype   & fp16, bf16, int8, etc. \\ \hline
    size\_bytes & uint64 \\ \hline
    \end{tabular}
    }
    \label{tab:chakra-tensor-schema}
\end{minipage}
\hfill
\begin{minipage}{0.21\textwidth}
    \centering
    \caption{Storage schema.}
    \resizebox{\linewidth}{!}{
    \begin{tabular}{|c|c|}
    \hline
    \textbf{Field Name} & \textbf{Data Type} \\ \hline
    id      & uint64 \\ \hline
    size\_bytes & uint64 \\ \hline
    device  & string \\ \hline
    \end{tabular}
    }
    \label{tab:chakra-tensor-storage-schema}
\end{minipage}
\vspace{-2em}
\end{table}

Tensor descriptors capture the structure and placement of data. 
Each tensor has a unique identifier, shape, and data type, with 
the \texttt{size\_bytes} field recording its storage footprint. The \texttt{storage\_id} field records the physical memory on which this tensor is allocated with \texttt{storage\_offset} to define the offset in that physical memory. Splitting tensor and its storage makes it possible to support tensor alias, for example, two tensors share the same storage but with different shapes. These descriptors allow compute and communication nodes to be linked consistently through tensor references. Tensor storage node represents a unique physical memory, it has a unique id, the memory size, and the device the storage resides.
This design satisfies the requirements mentioned earlier \autoref{sec:chakra-schema-design-requirements}. It is \emph{minimal yet extensible}, containing only core fields while supporting rich metadata through attributes. It is \emph{expressive}, capturing computation, memory, and collective or point-to-point communication. Finally, it is \emph{portable}, accommodating traces collected at different abstraction levels, and consumable by diverse downstream tools.

\textbf{Trace Storage.} 
In the current implementation, each NPU maintains its own ET. For a workload with $N$ NPUs, $N$ distinct traces are generated. This design simplifies collection and replay since each NPU records only its local activity without referencing the execution of others. However, this per-device isolation reduces the opportunity for global scheduling, where work could be reassigned or balanced across NPUs. The Chakra schema does not prohibit such an approach, but its default mode of operation assumes per-device traces.

\textbf{Trace Format.} 
Chakra ETs started with the Google Protobuf format for size and efficiency considerations.
Later, AMD introduced representing Chakra ETs in JSON format for better manageability (including human readability) and updated the ASTRA-sim simulator to support both formats of Chakra ETs to demonstrate the downstream tools can natively support both representations.
Today, Chakra traces can be collected and used in both JSON and Protobuf.

\vspace{-2mm}
\section{Chakra Trace Collection}
\label{sec:trace_collection}
\vspace{-1mm}

With the schema defined, the next logical step is to create traces. The Chakra ecosystem provides open-source tools for this, including a trace converter for post-processing native framework formats and a test case generator for synthetic trace creation. This tooling is complemented by native support in frameworks like PyTorch and NVIDIA NeMo, providing a direct path for trace generation. These tools and integrations support trace collection at two primary stages of the execution stack: pre-execution and post-execution, as shown in \autoref{fig:chakra_arch}.

Pre-execution refers to a stage where an ML model has not yet been explicitly optimized or tied to specific system configurations, such as compute engine type, memory bandwidth, or network topology. Collecting a pre-execution trace allows for performance projection across various systems (current and future), as it is not tethered to a particular configuration. Conversely, the post-execution stage involves optimizing and executing the ML model on an actual system. These traces accurately represent and capture real-world phenomena. They are valuable for replaying of compute/comms for tuning runtime libraries. 
However, these traces are tightly linked to the real system on which the model is run, with the  parallelization strategy and several optimizations baked in.
We discuss both of these in this section.




\insertFigure{dataflow1}{Chakra Trace Generation Flow from PyTorch}{0.98}{-1em}{-1em}
\insertFigure{Dependency}{Converted Trace example.}{0.96}{-1em}{-1.5em}

\vspace{-1mm}
\subsection{Post-Execution Traces}\label{sec:post-exec-trace}
\vspace{-1mm}


To collect post-execution traces, we leverage PyTorch’s profiling stack with two complementary tools: execution graph observer~\cite{pytorch_eg_observer} and Kineto~\cite{pytorch_kineto}. 
The execution graph observer records central processing unit~(CPU)~(i.e., host-side) execution, capturing the logical sequence of operator invocations and control dependencies that describe how the CPU launches graphics processing unit~(GPU) kernels. This forms the control-flow backbone of the workload.
Kineto, in contrast, profiles device-side activity on GPUs (or NPUs), recording kernel launches, durations, and stream identifiers with microsecond precision. It provides the fine-grained timing information that the observer lacks. Both tools can be enabled through standard PyTorch application programming interfaces~(APIs) or NVIDIA NeMo~\cite{nvidia_nemo} (see \texttt{NeMo/nemo/core/classes/modelPT.py}) without modifying user models. Furthermore, to capture modern LLM-era inference deployments, we extend Chakra's trace collection capabilities into serving frameworks like vLLM \cite{vllm_github}
. Specifically, we integrate the collection of Chakra traces directly into vLLM's native profiling system (see \texttt{vllm/profiler/wrapper.py}).



\autoref{fig:dataflow1} summarizes the Chakra trace generation flow.
\circled{1}~Running a PyTorch model produces a Chakra host ET (observer). Additionally, PyTorch can generate a profile trace (Kineto) that captures timing and device operator information. The Chakra host ET is sufficient for replay use cases since the trace is re-executed on a real system.
\circled{2}~The Chakra trace linker merges them into a unified dependency graph, resolving control, data, and synchronization edges.
\circled{3}~The Chakra trace converter validates the merged graph and emits a standardized Chakra ET following the Chakra schema.
\circled{4}~Downstream tools such as the ET feeder, simulators, and visualizers then consume the ET for analysis, and simulation. 
Each is described in detail in the subsequent sections.

\vspace{-1mm}
\subsubsection{Chakra Trace Linker}
\label{sec:chakra-trace-linker}
\vspace{-1mm}

The \emph{Chakra trace linker} merges host-side (CPU) and device-side (GPU or NPU) execution traces into a unified representation. Collected independently, these traces are otherwise disjoint, limiting holistic reasoning about control flow and performance. The linker resolves this gap by establishing explicit cross-domain dependencies and producing a single dependency graph in Chakra schema suitable for downstream tools. An example is shown in \autoref{fig:Dependency}.

\textbf{Why linking is necessary.}
Host traces collected through PyTorch’s observer contain CPU activity and call stack information, while Kineto traces provide fine-grained timing information for kernels and operators. However, Kineto does not encode dependency information, and PyTorch observer lacks detailed device timing. Neither source alone is sufficient to construct a comprehensive dependency graph. To address this gap, Chakra combines both sources: CPU information from the host trace and timing from Kineto. We also contributed a patch to PyTorch so that observer and Kineto traces share common identifiers, allowing events to be matched across the two sources; this pull request has been merged upstream. Still, merging raw traces is not enough. In order to build a complete dependency graph suitable for analysis and simulation, we explicitly reconstruct multiple classes of dependencies—control, data, and synchronization—on top of the raw trace data. Concepts such as inter-stream and intra-stream dependencies, also discussed in systems like Lumos \cite{liang2025lumos}, are modeled in Chakra and integrated into the unified representation.

\textbf{Control dependency.}
A control dependency indicates that the execution of one operation must follow another. 
For example, if function \texttt{A} calls function \texttt{B}, then \texttt{B} is control-dependent on \texttt{A}. 
Likewise, the return from \texttt{B} back to \texttt{A} also represents a dependency, since \texttt{A} cannot continue until \texttt{B} has completed. 
The linker reconstructs such causal ordering between host and device. 
This includes CPU function calls that launch kernels (CPU $\rightarrow$ GPU edges), 
returns or completion notifications from the device back to the host (GPU $\rightarrow$ CPU edges), 
and the ordering among host calls in the CPU call stack. 
These reconstructed links form the control-flow backbone of the unified trace and ensure that asynchronous operations are represented in the correct order.

\textbf{Data dependency.}
Producer-consumer relationships are established by matching tensor identifiers and buffer handles across host and device. For example, when the CPU produces a tensor later consumed by a device kernel, or when a device kernel generates an output read by the host, the linker inserts explicit data edges. 
Similarly, two GPU kernels have producer-consumer relationship such as compute operation followed by communication operation (inter-stream dependency), compute operation followed by another compute operation  (intra-stream dependency), or hierarchical collective communication (inter-stream dependency) linker connects those with data edges. 
This guarantees that data movement and its dependent computation are represented consistently.

\textbf{Synchronization dependency.}
The linker also inserts edges that capture ordering imposed by synchronization operations. This includes global synchronizations such as \texttt{cudaDeviceSynchronize()}, stream-specific calls like \texttt{cudaStreamSynchronize()}, and event-based coordination through \texttt{cudaEventRecord()} and \texttt{cudaStreamWaitEvent()}. Inspired by Holistic Trace Analysis~\cite{hta_github}, these synchronization edges are treated as additional dependency constraints that extend the unified graph. 
By encoding them, the linker enables critical-path analysis and ensures that both explicit and implicit synchronization points are visible to downstream tools. 


\subsubsection{Chakra Trace Converter}
\label{sec:execution-trace-converter}
\vspace{-1mm}


The Converter operates after the linker and has two goals: (1) verify the dependencies produced by linking, and (2) emit a standardized Chakra ET graph.

\textbf{Dependency verification.}
The Converter checks the linked graph for structural soundness and removes artifacts that hinder analysis. It enforces acyclicity via standard topological validation, prunes false or redundant edges (e.g., edges contradicted by per-stream order or duplicating implied relations), reconciles inter- and intra-stream constraints into a consistent ordering, and validates process group and domain consistency for communication nodes. All surviving edges are normalized into a single edge set with a \texttt{dep\_type} label (control/data/sync) and de-duplicated for determinism.

\textbf{Graph emission.}
Verified events are serialized into the Chakra schema as typed nodes (\texttt{COMP}, \texttt{MEM\_LOAD/MEM\_STORE}, \texttt{COMM\_SEND/RECV/COLL}) with optional timing hints (\texttt{start\_time\_micros}, \texttt{duration\_micros}) and extensible attributes (tensor sizes, ranks, group identifiers, tags, input-output~(IO) metadata). Edges are emitted as a cycle-free, canonical adjacency with stable ordering. The result is a single, standardized Chakra DAG.

\vspace{-1mm}
\subsection{Pre-Execution Traces}\label{sec:pre-exec-trace}
\vspace{-1mm}
Existing work has explored synthetic or compiler approaches for pre-execution trace generation. 
\textbf{STAGE}~\cite{stage} constructs symbolic tensor graphs 
enabling scalable synthesis of LLM benchmarks without relying on a specific runtime environment. 
\textbf{SimAI}~\cite{simai} has a custom domain specific language~(DSL) to describe LLMs and create a workload graph (that we have successfully ported to the Chakra schema)
with a focus on detailed kernel-level modeling of computation and communication, enabling more accurate runtime-specific performance analysis. 
\textbf{AMD} leverages an in-house analytical modeling framework to evaluate the performance of end-to-end AI application execution and generates pre-execution Chakra graphs.
\textbf{Flint}~\cite{flint} captures PyTorch compiler's Intermediate Representations (\texttt{torch.fx}) and converts them into Chakra graphs.
\textbf{LayerDAG}~\cite{li2024layerdag} leverages generative AI models to synthesize execution traces in order to obfuscate intellectual property while retaining representative performance characteristics.

These approaches highlight complementary motivations for different trace synthesizers: STAGE emphasizes scalability and coverage of large (and hypothetical) models, SimAI and AMD tool emphasize accuracy for specific platforms, Flint allows users to easily obtain high fidelity graphs from model source code as-is. LayerDAG emphasizes protecting intellectual property from trace release. Chakra's schema accommodates for all these pre-execution traces for design space exploration.


\vspace{-2mm}
\section{Chakra Downstream Use Cases}
\label{sec:use-cases}
\vspace{-1mm}

Inspired by the co-design stages shown earlier in \autoref{fig:codesign}, the Chakra ecosystem provides tooling and harnesses for three broad tasks: analysis, replay and simulation/emulation that are required at different times within the development cycle of AI platforms. The open schema enables interoperability across different stages and diverse open/proprietary tools.

\vspace{-1mm}
\subsection{Trace Analysis}
\label{sec:trace-analysis}
\vspace{-1mm}



\insertFigure{execution_graph_visualizer}{Chakra ET visualization example.}{0.98}{-1em}{-1.5em}

Chakra offers a range of open-source tools to help users visualize, analyze, and consume execution traces.
We describe these here briefly.
All these tools are available open-source at the MLCommons Chakra GitHub repository
and are under active development.

\textbf{Trace Visualizer.} The trace visualizer is designed to visualize execution traces in the Chakra schema---both the dependencies and the timeline.
An example of a visualization from the execution trace visualizer is presented in \autoref{fig:execution_graph_visualizer}.
The visualizer is a helpful tool for researchers to understand the structure of execution traces, or debugging.
By default, the visualizer encodes the names of tasks and dependencies between tasks. 
Users can easily modify the visualizer to encode additional metadata such as compute time and communication size. 

\textbf{Trace Reconstructor.}
A trace reconstruction tool consumes a Chakra ET graph and executes a policy-agnostic topological schedule (Kahn-style ready queue), which can be used for validation, benchmarking or visualization.


\textbf{Dependency-Aware ET Feeder}
The feeder ingests execution traces as a dependency graph and streams nodes to a simulator while strictly preserving the partial order defined by control and data edges. To bound memory usage, nodes are read in windows rather than loading the entire trace. When a node refers to a parent that has not yet appeared, it is placed into an unresolved set until the parent arrives. The feeder then elastically extends the window to resolve cross-boundary dependencies, ensuring correctness without preloading the full trace.
Each node maintains a count of unresolved predecessors. A node is ready once this count reaches zero and is inserted into a ready queue. The queue itself is policy-driven: nodes may be prioritized by measured start time, communication priority, or simply issued in first-in-first-out~(FIFO) order. Crucially, policies only arbitrate among ready nodes and therefore cannot violate dependency invariants. When a node completes, the feeder decrements the predecessor counts of its children, potentially unlocking new ready nodes. This process ensures that the emission order respects the original dependency structure.

With this design, correctness is guaranteed by construction, while scheduling policy remains pluggable. The approach yields deterministic emission orders under a fixed policy, scales linearly with the size of the trace, and keeps memory usage proportional to the window size rather than the entire input. As a result, the feeder can support a variety of backends—compute simulators while preserving the causal structure of the original execution trace.

\vspace{-1mm}
\subsection{Chakra Trace Replay on Current Systems}
\vspace{-1mm}

Benchmarking for ML systems has traditionally relied on carefully chosen applications that are adapted or simplified to expose specific behaviors. This process is time-consuming and difficult to maintain as workloads evolve.

Chakra replay takes an alternative path by relying on AI framework backend, e.g. PyTorch Aten and c10d backends, to re-execute compute and communication operations in ETs at scale on real systems. By using traces from production, collected with low-overheads, we can quickly generate portable benchmarks that can be used early stage platform evaluation,
subtrace replay, and scaled-down performance testing. This approach achieves high fidelity with production workloads as demonstrated for communication operations ~\cite{param_github} as well as full workloads ~\cite{liang2023mystique}. 

MLCommons is considering defining networking benchmarks using this methodology as well as improving existing MLPerf Storage benchmarks~\cite{mlperf-storage} to use Chakra replay. 

\subsubsection{Chakra Replay Benefits}
\vspace{-1mm}
Chakra replay offers several notable advantages in comparison with re-running the original workload.

\noindent
\textbf{Data Privacy and Accessibility.}
The replay process substitutes randomized input data wherever feasible to mitigate data-privacy concerns inherent in production environments. This design choice allows researchers and performance engineers to analyze workload behavior without requiring access to sensitive user data, which is often restricted.

\noindent
\textbf{Accelerated Debugging and Analysis.}
A Chakra trace encapsulates only the operations executed on the target devices. Consequently, the replay tool bypasses the model initialization and compilation phases—such as the potentially time-consuming PyTorch model compilation step—thereby substantially reducing turnaround time during debugging and performance analysis.

\noindent
\textbf{Fine-Grained Replay Control.}
The tool allows users to replay a selected subsequence of operators rather than the full trace. This capability facilitates targeted examination of specific computational regions for debugging or performance optimization.

\noindent
\textbf{Device Agnosticism.}
Because Chakra traces record operators at the ML framework level, they are typically device-independent. As long as the recorded operators are supported on the target hardware (for example, communication collectives), the replay can be executed on devices different from those used during the original trace collection.
\subsubsection{Chakra Replay Workflow}
\vspace{-1mm}
The replay tool operates through the following stages.

\noindent
\textbf{Process Initialization.}
A dedicated process is instantiated for each rank and corresponding device. Each process loads the Chakra trace associated with its rank.

\noindent
\textbf{Trace Parsing.}
The trace is parsed to identify the set of nodes to be executed according to the selected replay configuration—compute-only, communication-only, or full replay—and to determine whether to execute all operators or a user-specified subset.

\noindent
\textbf{Operator Initialization.}
The relevant operators are instantiated and initialized through the corresponding machine-learning framework interface.

\noindent
\textbf{Tensor Allocation.}
Input and output tensors are analyzed, and memory is provisioned according to the chosen tensor allocation strategy:

Pre-allocation mode allocates all input tensors before execution, trading increased memory consumption for lower allocation overhead.
Lazy allocation mode allocates tensors on demand and releases them when they leave scope, improving memory efficiency at the expense of additional allocation overhead.

\noindent
\textbf{Execution and Profiling.}
The operators are executed in the original recorded order. When profiling is enabled, the tool produces a detailed performance report, including kernel-level timing and execution statistics.
\noindent

\subsubsection{Collectives Accuracy Comparison}
\vspace{-1mm}
One of the challenges in AI model training is maintaining consistent model convergence across various device accelerators.
The new accuracy checker feature in Chakra replay addresses this by comparing the outputs of collective reductions. This is achieved by replaying the input collective tensors on different accelerators.

The accuracy comparison capability within the Chakra replayer facilitates the following:
\vspace{-1mm}
\squishlist
\item Validation of the relative difference of collective reduction outputs between diverse accelerator hardware and reduction algorithms.

\item Comparison of the precision loss with different datatypes both on the same accelerator type or across different accelerators types.

\item Debugging model collectives by replaying the model's input tensors to analyze the reduction behavior on the accelerator.
\squishend

Specific performance results are not presented here due to proprietiary nature of analysis. 


\subsection{Performance Projection for Future Systems}
\vspace{-1mm}

\subsubsection{What-if Analysis with Simulators}

Traditional compute or network simulations often rely on synthetic or overly simplistic workload models that fail to capture the nuanced complexity and dynamism of AI applications. As AI workloads increasingly dominate datacenter traffic, network research, in particular, must reflect their unique characteristics, including highly variable traffic patterns, burstiness, latency sensitivity, and high-bandwidth demands.
To this end, Chakra traces serve as workload specifications for forward-looking system studies. Here, the trace supplies operator types, tensor shapes, and dependency structures, while simulators or performance models can provide execution time breakdowns. In particular, given the coarse-grained compute adn communication operators in the Chakra schema, simulators can also be used to study software optimizations for compute and collective algorithms, which is not feasible when running over a fixed vendor software stack (e.g., CUDA and NCCL from NVIDIA or RCCL\texttrademark{} from AMD).

ASTRA-sim~\cite{astrasim-web} was one of the first open-source simulators to adopt Chakra, leveraging Chakra's ET feeder (\autoref{sec:trace-analysis}) to replace its custom workload format. This has enabled several co-design studies studying novel platforms---spanning new fabric topologies~\cite{libra}, wafer-scale systems~\cite{rashidi2025fred}, and topology-aware collective synthesis~\cite{tacos}. 
Today, Chakra trace support has also been enabled within proprietary simulators, e.g., by Scala Computing~\cite{scp}.



\subsubsection{Hardware-in-the-Loop Validation with Emulators}
\label{sec:hw-in-the-loop-use-case}

A key industry trend for accelerating multi-year AI supercomputer development is to \emph{shift left}, moving validation from full-scale clusters to earlier, component-level validation~\cite{hil_leftshift, hil_clusterless_ocp}. This hardware-in-the-loop (HIL) validation relies on emulators driven by high-fidelity workloads. Here, the Chakra ET serves as the portable workload specification. An emulator replays the trace to generate at-scale network dynamics, directing this stimulus at a physical device under test (DUT) like a network interface controller~(NIC) or switch. This process allows engineers to verify performance against simulation predictions and uncover implementation bugs that real-world traffic patterns expose. The viability of this trace-driven approach is demonstrated by its native adoption in open-source tools like Genie~\cite{genie} as well as commercial system emulators like Keysight AI Data Center Builder~\cite{keysight_kai_dc_builder}, which now support the Chakra format for workload import and replay.

\begin{table*}[t]
  \caption{Counts of key operations per GPU for a single epoch.}
  \centering
  \resizebox{2\columnwidth}{!}{%
  \begin{tabular}{|l|c|l|cccc|c|cccc|}
    \hline
    \textbf{Model} & \textbf{GPUs} & \textbf{Parallelization} &
      \multicolumn{4}{c|}{\textbf{Computation}} &
      \multicolumn{5}{c|}{\textbf{Communication}} \\
    \cline{4-7}\cline{8-12}
    & & & GeMM & Attn & ElemWise & Others & P2P & AllReduce & All2All & AllGather & ReduceScatter\\
    \hline
    \multirow{3}{*}{\makecell[l]{GPT3 5B \cite{gpt3}}} 
      & 8  & TP=8, w/ SP             & 37,248 & 6,144 & 24,832 & 165,207 & 0   & 514   & 0   & 18,816 & 12,416 \\
      & 8  & PP=8                     & 4,608  & 768   & 3,072  & 21,475  & 136 & 2     & 0   & 0      & 0     \\
      & 8  & FSDP=8                   & 4,656  & 768   & 3,104  & 9,400   & 0   & 17    & 0   & 784    & 400    \\
    \hline
    \multirow{2}{*}{\makecell[l]{GPT3 175B \cite{gpt3}}} 
      & 32 & TP=32 w/ SP              & 36,960 & 6,144 & 24,640 & 88,573  & 0   & 130   & 0   & 18,528 & 12,320  \\
      & 64 & TP=4, DP=2, PP=8, w/SP   & 9,216  & 1,536 & 3,072  & 26,615  & 67  & 67    & 0   & 4,683  & 3,078   \\
    \hline
    \multirow{3}{*}{\makecell[l]{Llama3 70B \cite{llama3}}} 
      & 8  & TP=4, PP=2               & 24,576 & 4,096 & 12,288 & 46,080  & 130 & 2     & 0   & 12,416 & 8,320  \\
      & 8  & TP=8                     & 49,536 & 8,192 & 24,832 & 85,634  & 0   & 16,897& 0   & 0      & 0     \\
      & 16 & TP=4, PP=2, DP=2         & 12,288 & 2,048 & 6,144  & 23,054  & 65  & 4,161 & 0   & 16     & 16    \\
    \hline
    \multirow{2}{*}{\makecell[l]{Mixtral 8x7B \cite{jiang2024mixtral}}} 
    & 8 & TP=2, EP=4               & 5912  & 1024   & 7751   & 16657  & 0  & 291     & 1024  & 1322      & 1122     \\
      & 32 & EP=8, PP=4               & 1,920  & 256   & 512    & 11,028  & 16  & 2     & 512 & 9      & 9     \\
      & 128& TP=4, EP=8, PP=4         & 1,920  & 256   & 512    & 10,801  & 16  & 131   & 512 & 659    & 531  \\
    \hline
    \multirow{1}{*}{\makecell[l]{Mixtral 8x22B\cite{jiang2024mixtral}}} 
      & 32 & EP=8, TP=4               & 3,596  & 896   & 6,333    & 13,277  & 0  & 243     & 896 & 1,134      & 905      \\
    \hline
    \multirow{1}{*}{\makecell[l]{DeepSeek-MoE \cite{deepseek-moe}}} 
      & 8  & EP=8                     & 27,456 & 896   & 6,784  & 46,938  & 0   & 18    & 1,728 & 23   & 23    \\
      \hline
      \multirow{1}{*}{\makecell[l]{DLRM \cite{DLRM}}} 
      & 8  & MP=8                     & 77 & 0   & 152  & 878  & 5   & 9   & 10 & 0   &  0     \\
      \hline
        \multirow{1}{*}{\makecell[l]{Resnet50 \cite{resnet}}} 
      & 2  & MP=2                     & 33 & 0   & 347  & 3263  & 0   & 15   & 0 & 0  & 0  \\
    \hline
  \end{tabular}
  }
  \label{tab:key_operations_comparison}
\end{table*}

     

\vspace{-2mm}
\section{Evaluation}
\label{sec:eval}
\vspace{-1mm}

We present a suite of case studies using a mix of open and commercial tools that support Chakra today to demonstrate the value of the ecosystem to the community.

\subsection{Chakra Trace Analysis}
\vspace{-1mm}


\insertFigure{runtime_idle_label}{Normalized execution time breakdown across workloads for traces collected on the system mentioned in~\autoref{sec:eval}. For each workload, we show measured performance from Kineto (left) and the performance via trace reconstruction through Chakra (right).}{0.98}{-1em}{-1em}

\insertFigure{coll_comm_ib_perf}{Total collective communication runtime comparison at 400\,Gb/s and 100\,Gb/s InfiniBand.
Measured on training Mixtral-8×22B with 32~GPUs (four HGX-8×H200 nodes, TP/SP=4, EP=8) and the global batch size of 32.}{0.98}{-1em}{-1em}




\noindent
\textbf{Evaluation System.}\label{sec:evaluation_sys}
We collect traces from the Georgia Tech AI Makerspace~\cite{makerspace}, a supercomputer hub providing a diverse set of largescale compute resources to students and researchers. Here we use 128 NVIDIA H100/200 GPUs interconnected with Infiniband HDR100 (100\,Gb/s), dual 32-core Intel Sapphire Rapids CPUs, and DDR5 DRAM. We also collect traces from Hewlett Packard Enterprise, across 32 NVIDIA H200 GPUs interconnected with Infiniband HDR400, dual 32-core Intel Xeon 8562Y CPUs, and DDR5 DRAM. Our software environment includes PyTorch 2.5, NVIDIA NeMo 24.07, and Megatron 0.10.0. 
We track control and data flow by analyzing the profiling results in both trace view and graph execution format. The details of the traces collected for this study are presented in \autoref{tab:key_operations_comparison}. We have made these traces publicly available online.\footnote{https://github.com/mlcommons/chakra/wiki/Chakra-Trace-Library}

\circled{1} \textbf{Runtime Analysis.}
\autoref{fig:runtime_idle_label} shows each workload evaulated under comparing the Kineto and Chakra traces. Computation (red) and exposed communication (blue) in Kineto and Chakra align closely. However, for co-design analysis, Chakra excludes the idle time (gray) that occurs between different GPU and CPU kernels.

\autoref{fig:coll_comm_ib_perf} presents the Chakra trace analysis results for the Mixtral-8×22B model executed with identical tensor and expert parallelism configurations across InfiniBand networks of different bandwidths. Because the scale-up interconnect is confined to eight GPUs per node, it is primarily utilized for tensor parallelism~(TP), particularly for non-mixture-of-experts~(MoE) components, while most expert related communications occurs over the slower scale-out InfiniBand network. Using Chakra for analyzing total duration for different collective communications, we observe that a 4.0$\times$ slower InfiniBand bandwidth leads to an approximately 4.1× and 4.4× slowdown in All-to-All and All-Gather, respectively. Also, higher-bandwidth InfiniBand also exhibits lower latency, so the effective slowdown is less than the theoretical 4× ratio and this effect becomes more prominent in the All-Reduce kernel, which has substantially smaller communication volume.

Additionally, by encoding memory utilization collected from Kineto for compute nodes, Chakra tracks the total utilization at different timestamps throughout the training epochs. \autoref{fig:output_memory_track} illustrates the memory usage contributed by various compute nodes during the execution.





\insertFigure{output_memory_track}{GPU memory utilization for different LLM models during one training step. Traces are aligned relative to the start of each epoch. Each model and its corresponding parallelization match the first entry (row) in \autoref{tab:key_operations_comparison} using the same number of GPUs.}{0.98}{-1em}{-1em}

\insertSubFigs
  {mixtral-8x22NodeDurations-new.pdf}     
  {Cumulative distribution function of node durations.}  
  {compute_duration}                      
  {mixtral-8x22ComputeDataDependencies-new.pdf}  
  {Distribution of data dependencies.}    
  {compute_deps}                          
  {Compute characteristics of the Mixtral-8x22-Chakra trace. 
   (a) Most compute kernels complete within $2$--$10^2$\,µs. 
   (b) The majority of nodes have $10$--$500$ parent data dependencies.}  
  {compute_analysis}                      
  {-1em}  

\circled{2} \textbf{Kernel Analysis.}
Breaking down end-to-end computation and communication into individual kernel nodes, Chakra extracts precise operation counts and timings across diverse models, parallelization strategies, and configurations, and encodes them in a consistent graph representation. \autoref{tab:key_operations_comparison} summarizes the raw operation counts across various models, while \autoref{fig:compute_duration} and~\autoref{fig:compute_deps} visualize the cumulative distribution function~(CDF) of compute durations and data dependencies obtained from the exemplified Mixtral-8x22B trace. 

\subsection{Trace Replay Case Studies}
\vspace{-1mm}
Replaying Chakra ETs on real systems allows reproducing the exact workload behavior either fully (replay both compute and comms operations) or partial replay (replay selective operations). The latter enables users isolate specific system components from overall workload execution to identify performance optimization opportunities (e.g. focus on networking improvements). 

We present a case study for replaying communication operations for a Megatron-LM 43B GPT model (48 transformer layers) on four H100 nodes (32 ranks) with pipeline parallelism~(PP) factor of four, TP factor of four, and data parallelism~(DP) factor of two. \autoref{tab:chakra-replay-gpt3} shows the NCCL kernel bus bandwidth report from Chakra communication replay. The NCCL kernel bandwidth is close but typically faster than the original kernel in the baseline run due to lack of memory contention on overlapping with compute. 

The replay methodology helped a vendor identify a set of communication operations (AllGather and ReduceScatter) across microbatches that could be run in parallel leading to a 2$\times$ improvement in communication collective performance.


\begin{table}[t]
\centering
\footnotesize
\setlength{\tabcolsep}{3pt}
\caption{NCCL kernel bandwidth report from Chakra replay for top 10  NCCL kernels by message size, across different process group sizes.}
\begin{tabular}{lrrrrr}
  \toprule
  \textbf{Kernel} & \textbf{Size} & \textbf{Rks} & \textbf{Dur} & \textbf{BW Base} & \textbf{BW Ratio} \\
   &  &  & (ms) & (GB/s) &  \\
  \midrule
  ReduceScatter f32 & 9.0G   & 2 & 15.790 & 243.6 & 1.261 \\
  ReduceScatter f32 & 9.0G   & 2 & 57.378 & 157.7 & 1.106 \\
  AllGather         & 4.5G   & 2 & 7.848  & 289.2 & 1.065 \\
  AllGather         & 4.5G   & 2 & 7.491  & 312.8 & 1.034 \\
  ReduceScatter f32 & 404M   & 2 & 22.428 & 153.9 & 1.214 \\
  ReduceScatter f32 & 320.1M & 2 & 0.778  & 174.6 & 1.236 \\
  ReduceScatter f32 & 256M   & 2 & 3.887  & 177.5 & 1.013 \\
  AllGather         & 202M   & 2 & 1.264  & 209.0 & 0.784 \\
  ReduceScatter f32 & 192.1M & 2 & 0.494  & 170.3 & 1.197 \\
  AllGather         & 160.0M & 2 & 0.395  & 164.6 & 1.291 \\
  \bottomrule
  \end{tabular}
  \vspace{-1mm}
  \label{tab:chakra-replay-gpt3}
  \end{table}

\subsection{Hardware-in-the-Loop Emulation Case Study}
\vspace{-1mm}

We present a case study 
demonstrating the HIL validation methodology described in \autoref{sec:hw-in-the-loop-use-case}. The experiment investigates the performance of a physical Ethernet fabric under a realistic MoE workload to uncover emergent, system-level bottlenecks related to congestion control.

\noindent
\textbf{Emulation System.}
The physical system under test (SUT) consisted of a four 12.8T switch fabric arranged in a 1:1 Clos topology at 400\,Gbps port speed. The fabric was configured with standard data center quantized congestion notification~(DCQCN) for congestion control. Realistic AI traffic was generated using Keysight AI DCB, a system emulator capable of replaying Chakra ET to generate high-fidelity remote direct memory access~(RDMA) traffic~\cite{hw_in_loop_ocp}. This allows for injecting precise, workload-specific communication patterns into the physical SUT.

\noindent
\textbf{Target Workload.}
The experiment utilized a synthetic Chakra ET designed to model the communication patterns characteristic of a modern MoE training iteration. Unlike simpler models that rely primarily on a single type of collective, MoE workloads are defined by the frequent interleaving of both All-Reduce and All-to-All collective operations. These two collectives represent opposite extremes of communication patterns: All-Reduce operations typically involve a few high-bandwidth flows, while All-to-All operations create a mesh of many low-bandwidth flows. This mixed-collective pattern was injected into the SUT to analyze its performance under realistic contention.

\insertFigure{combined_busbw}{Bus bandwidth per iteration when (a) All-Reduce (b) All-to-All (c) mixing All-to-All and All-Reduce in one time span.}{0.93}{-1.5em}{-1em}

\insertFigure{hw-in-loop-cdf-ar-a2a-new}{Mixing collectives results of CDF.}{0.9}{-1em}{-2.3em}


\textbf{Result.}
The experiment revealed a significant performance anomaly when interleaving All-Reduce and All-to-All collectives.
While both collectives show stable performance in isolation (\autoref{fig:combined_busbw}(a) and (b)), the mixed workload in \autoref{fig:combined_busbw}(c) demonstrates a mutually detrimental interaction.
The most visible impact is on the All-Reduce operations. After being interleaved with an All-to-All collective, their performance becomes highly variable and fails to consistently achieve the near line-rate bandwidth seen in isolation. Concurrently, while the aggregate throughput of All-to-All appears stable, its constituent flows are negatively impacted. Upon investigating, we found that the high-rate All-Reduce flows trigger DCQCN's congestion control mechanism, which in turn disproportionately throttles the many small flows of the All-to-All, creating stragglers. 
\autoref{fig:hw-in-loop-cdf-ar-a2a-new} discusses the CDF for concurrent execution of two collective communicaitons (All-to-All and All-Reduce) extended from the isolated. We observe that the mixing incurs a long-tail \textit{flow} completion time distribution for All-to-All creating stragglers that increase overall \textit{job} completion time.

\insertFigure{what_if_network_bw_dgx}{Communication time for different network topology and bandwidth with Mixtral 8x7B target.}{1}{-2.5em}{-1em}

\subsection{Trace Simulation Case Studies}\label{sec:trace_simulation}
\vspace{-1mm}

Next, we demonstrate the inter-operatability Chakra by running it through two distinct simulators. The first is the open-source 
ASTRA-sim~\cite{astrasim-web} and the second is the commercial Scala Compute Platform (SCP).\footnote{SCP~\cite{scp} is a simulator built over ns-3 
to perform large-scale simulations in AWS Cloud.}

\subsubsection{Different Network Topology and BW using AstraSim}
\vspace{-1mm}

\textbf{Setup.} 
We assume a system with eight H100, and we vary the network topology between switch, ring, and fully-connected. Additionally, we test bandwidths ranging from 75\,GB/s to 900\,GB/s. The Mixtral 8×7B model serves as the workload for this evaluation. 

\textbf{Results.}
\autoref{fig:what_if_network_bw_dgx} illustrates the normalized communication time of workloads while varying the network bandwidth. We observe the following:
(1) Even with the same end-link bandwidth, different network topologies exhibit distinct communication times. In general, the switch topology achieves the best performance, followed by ring, and finally fully-connected. We suspect this is because the switch topology enables better link utilization, whereas other topologies contain links that are not actively used, especially in the fully-connected case.
(2) As the bandwidth increases, the communication time eventually converges and stops improving, since latency becomes the dominant factor compared to bandwidth. This observation suggests that to further accelerate communication, one may need to increase the chunk size to mitigate the impact of latency.

\subsubsection{Network Transport Analysis using SCP}
\vspace{-1mm}

\textbf{Setup.} We simulated a 256 A100 GPU system with fully-provisioned 51.2\,Tb/s ethernet switches in a two-tier Clos topology with 800\,Gb/s inter-switch links and 400\,Gb/s NICs, running RDMA over converged Ethernet~(RoCE) v2. We simulated a Llama-2 trace \cite{stg}, with a 64 DP $\times$ 2 TP $\times$ 2 PP $\times$ 1 SP parallelism.

\textbf{Results.} The distribution of NICs by transmit utilization is shown in the \autoref{fig:SCALA_RoCE_NIC_RX_TX}(a) heatmap.
Notably, after 300\,ms, a large fraction of NICs are no longer transmitting at high rates.
This arises from the synchronous nature of LLM workloads: prolonged intervals of low network utilization reflect GPUs spending substantial time on computation rather than communication (due to the nature of more DP than TP and PP in the experiment setup) or waiting to receive data from other GPUs.  Considering the receive-side NICs in \autoref{fig:SCALA_RoCE_NIC_RX_TX}(b), we observe oscillatory behavior driven by the workload structure. Even with a fully provisioned topology, the average link utilization remains well below the 400\,Gb/s line rate, with extended periods of underutilization while GPUs are computing or waiting on dependencies. 
In summary, Chakra provides an effective way to capture dependencies within and across GPUs, enabling fine-grained simulation of LLM workloads 
and enable studying the effect of network transport on end-to-end performance.


\insertFigure{SCALA_RoCE_NIC_RX_TX}{Results of RoCE v2 NIC transmit receive.}{0.98}{-1em}{-2em}

\subsection{Inference Trace Analysis}\label{sec:inference_analysis}

As introduced in ~\autoref{sec:post-exec-trace}, Chakra’s native trace collection capabilities were originally designed for training workloads in frameworks such as PyTorch and NVIDIA NeMo, and are now being extended to inference engines like vLLM to capture emerging and dynamic inference mechanisms not present in training. Here, we analyze traces collected through our ongoing vLLM integration across three key inference mechanisms. All experiments are conducted on vllm v1 \cite{vllm_pr30520} with supportance of prefill/decode disggregation. The same GPU/CPU platform is used as in \autoref{sec:evaluation_sys}, at a smaller scale than our training studies due to the nature of inference.
\subsubsection{MoE Token Routing} Unlike training frameworks that pad or drop tokens to maintain balanced All-to-All communication in NeMo, inference workloads preserves every token and creates a dynamic and load-imbalanced workload pattern that will be assigned across different experts being held on different GPUs. In Chakra, we embed the per-expert bin counts (e.g. [1, 2, 5, 0, 0, 0, 0, 4] for eight experts) to the MoE routing nodes. \autoref{fig:routing_prefill_single} shows the profiled results of MoE token routing among two GPUs, so each GPU rank will hold four experts and receive an imbalanced amount of tokens. 

\begin{table}[t]
\centering
\caption{Comparison of KV-cache offloading and baseline (no offloading) operations in Chakra collected traces for Llama3-8B.}
\label{tab:kv_offloading_comparison}
\scriptsize
\setlength{\tabcolsep}{2.5pt}

\begin{minipage}{0.97\columnwidth}
\centering
\begin{tabular}{lcccc}
\toprule
& \multicolumn{2}{c}{\textbf{Baseline}} & \multicolumn{2}{c}{\textbf{Offloading}} \\
\cmidrule(lr){2-3} \cmidrule(lr){4-5}
\textbf{Operation} & \textbf{Count} & \textbf{GPU Time (ms)} & \textbf{Count} & \textbf{GPU Time (ms)} \\
\midrule
Memcpy HtoD    & 2,334 & 2.088   & 4,857 & 11.400 \\
Memcpy DtoH     & 387   & 0.895   & 5,958 & 216.484 \\
\texttt{start\_load\_kv}  & N/A   & N/A     & 784   & 6.337 \\
\texttt{start\_store\_kv} & N/A   & N/A     & 776   & 215.068 \\
\bottomrule
\end{tabular}
\vspace{-1em}
\end{minipage}
\end{table}

\subsubsection{Memory Offloading between CPU and GPU}
To demonstrate Chakra’s ability to analyze memory offloading, we collect traces by forcing key-value~(KV) cache offloading under limited GPU memory. We then isolate trace nodes captured from the recorded function \texttt{start\_load\_kv} in vLLM’s KV connector. \autoref{tab:kv_offloading_comparison} quantifies the impact of offloading by tracking KV store/load events and the resulting extra host~(CPU)–device~(GPU) memory interactions, highlighting Chakra’s ability to expose system-level bottlenecks in memory-bound inference settings.

\insertFigure
  {routing_prefill_single}
  {Distribution of token routing among two expert parallel rank for each model layer. The input has six tokens and the model used is Mixtral 8x7B with 32 layers.}
  {0.95}
  {-1.5em}
  {-1.2em}

\insertFigure{kv_transfer_per_layer}{Runtime breakdown of the KV cache transfer for inferencing Llama3-8B between one prefill and decode GPU. The captured trace denotes the per-layer (32 layers for Llama3-8B) send and receive latency between two GPUs.}{0.95}{-1.5em}{-2em}

\subsubsection{KV-Cache Transfer} 
In inference, when disaggregating prefill and decode stages on different GPUs~\cite{splitwise, zhong2024distserve, MIST}, it introduces unique point-to-point communication requirements. To capture this, we profiled the message sizes of the point-to-point messages responsible for transferring the KV cache between dedicated prefill and decode GPUs as shown in \autoref{fig:kv_transfer_per_layer}. 

\section{Lessons Learned and Extensions}\label{sec:lessons_learned_extensions}

We discuss lessons learned while working with Chakra. We then point out how our lessons evolved into current and future extensions.

\subsection{Lessons Learned}\label{sec:lessons_learned}

\begin{itemize}[itemsep=2pt, topsep=2pt, parsep=0pt, partopsep=0pt]
\item A graph-based ET effectively bridges the gap between high-level frameworks (e.g. PyTorch), LLM frameworks (e.g. NeMo), and downstream replay/simulation tools; provide a common abstraction from multiple industry participants to exchange critical bottlenecks without revealing sensitive intellectual property~(IP).
\item By abstracting execution into compute, memory, and communication nodes with explicit dependencies, we can enable portable and reproducible benchmarking without disclosing model weights or datasets. This has allowed Meta/NVIDIA to identify bottlenecks in overlapping compute/comms that were invisible in isolated benchmarks \cite{saeed_enable_2021}.
\item While the original Chakra capture mechanism was mainly designed for eager mode operator execution, we continue to expand the trace to include information like synchronization dependencies, KV transfers and data loader nodes for MLPerf Storage benchmarks. This enables the trace to provide a higher fidelity represntation of the actual system behavior.
\item Our experience reaffirms a cyclic process from productive SW/HW codesign: observe (in production) $\rightarrow$ reproduce (via replay) $\rightarrow$ design/evaluate new solutions (via simulators) $\rightarrow$ deploy (in production). To summarize, Chakra acts as the standardized mechanism for representing key performance aspects of a workload through every stage of the co-design cycle.
\end{itemize}

\subsection{Ongoing Extensions}\label{sec:ongoing_extensions}
\subsubsection{Handling Large-Scale Traces}
\vspace{-1mm}

The Chakra framework prioritizes completeness and fidelity in trace representation, which is essential for accurate performance modeling and replay. For very large workloads, such as billion-parameter language models or training runs with millions of operations, this design choice results in execution traces that can span multiple gigabytes. While these trace sizes reflect the true complexity of modern ML workloads, they also present practical challenges for storage, exchange, and initial loading. This approach represents an intentional trade-off: preserving full operational detail enables high-fidelity simulation and faithful bottleneck analysis, which would be compromised by lossy compression or aggressive summarization. To improve usability at scale without sacrificing fidelity, future enhancements could include optional lossless compression for storage and transfer, alongside hierarchical indexing to enable partial loading and selective replay. Such optimizations would enhance the scalability of the framework while maintaining its core accuracy guarantees for large-scale analysis.

\subsubsection{Infrastructure Abstraction}
\vspace{-1mm}

While Chakra provides a common representation for workload execution traces, infrastructure descriptions still rely on tool-specific formats that limit portability and cross-platform comparison. Just as Chakra enables workload exchange without exposing proprietary model details, a standardized infrastructure abstraction would allow hardware configurations to be shared and evaluated systematically across teams and organizations.
Graph-based infrastructure representations that capture compute nodes, memory hierarchies, and interconnect topologies in a portable format are promising. Such abstractions naturally complement Chakra’s execution graph model, enabling infrastructure-aware performance projection and topology-sensitive scheduling. Paired with Chakra ETs, standardized infrastructure descriptions can further support cross-platform trace reuse, interconnect design comparisons, and feasibility analysis of parallelization strategies under realistic resource constraints.
InfraGraph~\cite{infragraph} is an emerging effort within the MLCommons Chakra working group along this direction, complementing Chakra with graph-based infrastructure modeling for the trace ecosystem.

\subsubsection{Chakra for MLPerf Storage}
\vspace{-1mm}

The MLPerf Storage benchmark ~\cite{mlperf-storage} measures how fast storage systems feed the input data to the training model. Its current implementation replaces all compute and communication with a fixed delay and only replays storage operations. This limits fine-grained replay of storage together with compute and communication, and reduces portability across systems. We are working with the MLPerf Storage working group to integrate storage operations into Chakra ETs and enable replay for MLPerf Storage benchmarking.

\vspace{-0.5em}
\section{Related Work}
\label{sec:related_work}
\vspace{-0.2em}

The technique of collecting traces to understand system internals and identify bottlenecks is widely adopted and deployed. Google, for instance, collected and released instruction traces from its servers for representative cloud workloads~\cite{google_workload_traces}. Chakra is a similar effort aimed at ML systems research, specifically for distributed ML systems. In Chakra, we proposed a common schema for ML execution traces. The GOAL format  is another post-execution trace that builds around the communication operations, and is used to study both ML and high-performance computing~(HPC) workloads~\cite{atlahs}
As ML tasks are often represented as graphs, numerous graph schemas have been proposed. ONNX is one of the most popular graph schemas, designed to facilitate the exchange of models between different ML frameworks. Although ONNX shares similarities with Chakra, such as enabling data exchange between different teams, the objectives of the two differ. ONNX focuses on the exchange of \textit{models} between various frameworks, while Chakra's primary goal is to exchange \textit{execution traces} between different teams. While their purposes differ, ONNX and Chakra can be complementary---models expressed in ONNX can naturally yield execution traces encoded in Chakra.
When collected at the pre-execution stage, traces closely resemble the model itself, with additional annotations. These traces are similar to the graphs used in prior autotuning studies to optimize ML model parallelization strategies~\cite{wang2019parallel, santhanam2021distir, schaarschmidt2021automap}. Unity is a representative example: it jointly applies algebraic transformations and parallelization, and introduces a parallel computation graph to support such optimization. In contrast, Chakra is not limited to pre-execution traces; it also supports post-execution traces that capture real system dynamics, such as compute–communication overlap and idle periods, enabling more faithful replay and simulation.    
\vspace{-2mm}
\section{Conclusion}
\label{sec:conclusion}
\vspace{-1mm}

In this paper, we present Chakra, an MLCommons-enabled ecosystem for benchmarking and co-designing current and future systems. At its core, Chakra defines an open graph schema, called execution traces, to specify distributed workloads and capture key operations and dependencies. We also develop a complementary set of tools and capabilities for collecting, analyzing, and adopting Chakra execution traces across a wide range of simulators, emulators, and benchmarks. By addressing ecosystem fragmentation and workload-sharing barriers that lead to siloed optimizations and longer time-to-market, Chakra enables a more open and reusable co-design workflow. As future work, we are actively improving trace-size scalability, AI platform infrastructure standardization, and storage extensions.

\vspace{-1mm}
\section*{Acknowledgements}
\vspace{-1mm}

We acknowledge Srinivas Sridharan, Taekyung Heo, Joongun Park, and Brian Coutinho for the initial prototyping of Chakra. We thank Matt Bergeron, Wenyin Fu, Zhaodong Wang, Shengbao Zheng from Meta who helped develop the tooling and vision for early versions of Chakra.
We sincerely thank MLCommons for their role in establishing the Chakra Working Group and the technical inputs. We appreciate the support for Chakra from a large set of community advocates: Shashi Gandham (Meta), Alex Bortok and Ram Periakaruppan (Keysight), Arindam Mallick and Debjyoti Bhattacharjee (IMEC), Chun Liu (ByteDance), Ulf Hanebutte (Marvell), Samantika Sury (HPE), Oana Balmau (McGill University). We appreciate Matthieu Bloch and Aaron Jezghani in helping us leverage the AI Makerspace within the College of Engineering (RRID:SCR\_028058) at the Georgia Institute of Technology, provided by the Partnership for an Advanced Computing Environment (PACE) (RRID:SCR\_027619), for collecting a library of traces for this effort.

The results and comparisons provided in this paper are meant to showcase the value of the Chakra ecosystem and are not a direct endorsement of any specific vendor's hardware or software products.
AMD, the AMD Arrow logo, Instinct, RCCL, and combinations thereof are trademarks of Advanced Micro Devices, Inc. PyTorch, the PyTorch logo and any related marks are 
trademarks of The Linux Foundation. TensorFlow, the TensorFlow logo and any related marks are trademarks of Google Inc. NCCL, NeMo and CUDA are trademarks of NVIDIA Corp. Other product names used in this publication are for identification purposes only and may be trademarks of their respective companies.


\nocite{langley00}

\bibliography{references}
\bibliographystyle{mlsys2026}

\clearpage
\appendix
\section{Artifact Appendix}

\subsection{Abstract}

The artifact contains the source code and instructions for Chakra trace generation, linking, conversion, and analysis. In particular, it includes:
(i) utilities to Chakra collected host-side and device-side traces collected from nemo, as shown in \autoref{sec:trace_collection}
(ii) the Chakra Trace Linker and Converter, which merge host and device traces into a unified dependency graph and emit a standardized Chakra Execution Trace (ET) \autoref{sec:chakra-trace-linker},
(iii) downstream use cases of Chakra traces for execution-trace analysis, including Astra-sim simulation and \autoref{sec:trace_simulation} (iv) plotting scripts to reproduce Figures \ref{fig:runtime_idle_label}, \ref{fig:coll_comm_ib_perf}, \ref{fig:output_memory_track}, and \ref{fig:what_if_network_bw_dgx}.

The artifact evaluates two representative workflows. First, it demonstrates trace linking and conversion using pre-collected traces from NVIDIA NeMo. Second, it demonstrates a downstream use case by using Chakra traces to drive Astra-sim simulations different communication topologies.

\subsection{Artifact check-list (meta-information)}

{\small
\begin{itemize}[itemsep=0pt, topsep=0pt, parsep=0pt, partopsep=0pt]
  \item {\bf Program:} Chakra Trace Linker/Converter, trace processing utilities, Astra-sim integration scripts, and plotting scripts
  \item {\bf Compilation:} Python package installation for Chakra; Astra-sim built in Docker container
  \item {\bf Model:} Mixtral 8$\times$7B trace used for simulation
  \item {\bf Data set:} Pre-collected NVIDIA NeMo traces (PyTorch Observer + Kineto)
  \item {\bf Run-time environment:} Linux; Docker; Python 3.10+
  \item {\bf Hardware:} CPU-only machine
  \item {\bf Metrics:} ET validity, dependency statistics, simulated runtime under varied communication topologies, and reproduced analysis figures
  \item {\bf Output:} Chakra ETs, Astra-sim results, and reproduced figures
  \item {\bf How much disk space required (approximately)?:} 10\,GB
  \item {\bf How much time is needed to prepare workflow (approximately)?:} around 2 hours to link chakra-enabled nemo traces and convert it to chakra et that can be consumed by downstream simulator. After conversion, it took around 10-20 minutes to run the Astra-sim simulation.
  \item {\bf Publicly available?: } Yes
  \item {\bf Code licenses (if publicly available)?:} Apache-2.0
\end{itemize}
}

\subsection{Description}

\subsubsection{How to access}

The artifact is publicly available both as a GitHub repository and as an archived Zenodo release. It includes the Chakra trace processing pipeline, example traces, Astra-sim configurations, and the scripts needed to reproduce the results reported in the paper. Detailed setup and usage instructions are provided in the README under the mlsys26 directory of the Chakra repository (\url{https://github.com/mlcommons/chakra/tree/mlsys26}) and in the corresponding Zenodo archive (\url{https://doi.org/10.5281/zenodo.19636323}).

\subsubsection{Hardware dependencies}

All evaluations can be executed on a CPU-only machine. The trace linking and conversion pipeline operates on pre-collected trace files, and the downstream Astra-sim experiments run entirely in simulation.

\subsubsection{Inputs}

The artifact includes pre-collected NVIDIA NeMo traces collected via Chakra-enabled PyTorch profiling (Execution Graph Observer for host and Kineto for device). These traces serve as the inputs to the Chakra trace linking and conversion pipeline. The artifact also includes processed traces and configuration files used for simulation and the details are in the README.

\subsubsection{Evaluation workflow}

The artifact evaluation consists of three steps:

\begin{itemize}[itemsep=0pt, topsep=0pt, parsep=0pt, partopsep=0pt]
  \item \textbf{Trace linking and conversion.}  
  The provided NeMo host and device traces are linked to construct a unified dependency graph, which is then converted into a standardized Chakra ET.

  \item \textbf{Downstream simulation.}  
  The generated Chakra ET is used as input to Astra-sim to simulate execution of Mixtral 8$\times$7B. By varying the communication topology, users can observe the corresponding differences in simulated runtime.

  \item \textbf{Figure reproduction.}  
  The plotting scripts operate on the processed traces and simulation outputs to reproduce Figures \ref{fig:runtime_idle_label}, \ref{fig:coll_comm_ib_perf}, \ref{fig:output_memory_track}, and \ref{fig:what_if_network_bw_dgx} from the paper.
\end{itemize}

\subsubsection{Expected results}

Successful execution of the artifact should produce:
\begin{itemize}[itemsep=0pt, topsep=0pt, parsep=0pt, partopsep=0pt]
  \item A valid Chakra ET successfully generated from the provided NeMo traces that can be used for various downstream analysis
  \item Astra-sim simulation logs under different communication topologies, used for Figure~\ref{fig:what_if_network_bw_dgx}
  \item Reproduced figures comparable to Figures~\ref{fig:runtime_idle_label}, \ref{fig:coll_comm_ib_perf}, \ref{fig:output_memory_track},
\end{itemize}

\clearpage



\end{document}